# Learning Perceptual Manifold of Fonts

Haoran Xie*    Yuki Fujita    Kazunori Miyata

Japan Advanced Institute of Science and Technology

xie@jaist.ac.jp

**Abstract**

Along the rapid development of deep learning techniques in generative models, it is becoming an urgent issue to combine machine intelligence with human intelligence to solve the practical applications. Motivated by this methodology, this work aims to adjust the machine generated character fonts with the effort of human workers in the perception study. Although numerous fonts are available online for public usage, it is difficult and challenging to generate and explore a font to meet the preferences for common users. To solve the specific issue, we propose the perceptual manifold of fonts to visualize the perceptual adjustment in the latent space of a generative model of fonts. In our framework, we adopt the variational autoencoder network for the font generation. Then, we conduct a perceptual study on the generated fonts from the multi-dimensional latent space of the generative model. After we obtained the distribution data of specific preferences, we utilize manifold learning approach to visualize the font distribution. In contrast to the conventional user interface in our user study, the proposed font-exploring user interface is efficient and helpful in the designated user preference.

## 1. Introduction

Generative models have been explored intensively nowadays, as the rapid development of deep learning applications in computer vision and computer graphics fields. The representative approaches include the neural network frameworks of variational autoencoders (VAEs) [1] and Generative Adversarial Networks (GANs) [2]. The emerging progress of the artificial intelligence techniques has promoted machine intelligence over human intelligence in various applications, such as Alpha Go. From the other side, it is well known that the generative models can create helpful creative operations to augment human intelligence [3]. However, it is still challenging to understand the human perception in the conventional learning processes. In this work, we try to combine the machine process and human process together to achieve the perceptual adjustment of learning results. As a specific target, we focus on the character font generation in the above approach, because the fonts play crucial role in the representation of our daily information. In recent times, a plenty of character fonts have been created and published online such as in the widely used Google Fonts, Font Squirrel. A user may choose the preferred types and shapes of fonts for special purposes. However, the fonts used in festival posters and business documents may be entirely different from each other. Usually, the choice of fonts requires the skill and experience of a professional designer as it is always difficult to choose the most suitable fonts from online sources or font lists in document files to meet the user's perception need.

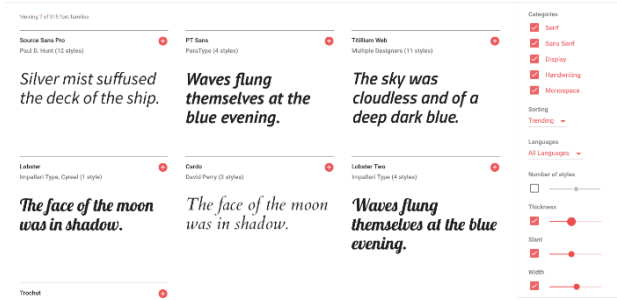

Figure1. Conventional font-exploring interface.

It is non-trivial for common users to explore and edit a specific font style to obtain the desired fonts from the font. Figure 1 shows a conventional font-exploring user interface of Google Fonts. In order to search for the desired fonts, the user is asked to select different image features of fonts. The image features may include font thickness, slant, and width. This type of exploring user interface is inefficient because human's font perceptions cannot easily be separated into specific features understanding the meaning of these tags may be challenging for non-experts. For example, it is difficult to clarify which font features are the most appropriate for creating an appealing festival poster. The same issues also arise in Font Squirrel where tags are used to display the font features. Recent research presents the font manifold in two-dimension (2D) by learning the nonlinear mapping from existing fonts [4]. In this previous work, a common user can easily explore fonts using the manifold with smooth interpolations. However, it is still challenging to generate and explore the fonts with specific perception by adjusting geometrical font features.

To address the above issues, we try to combine both machine intelligence and human intelligence in the process, where machine can generate numerous font images and human workers adjust the perception choice for the generated results. Especially, we propose a perceptual font manifold with different perception purposes. In contrast to the font matching approach with energy-based optimization process, which is time-consuming and difficult to implement, we directly applied the latent space from the generative model to construct the font manifold thanks to the rapid development of generative models based on deep learning approaches. In contrast to the previous work [3], the approach proposed in this paper is more straightforward and easier to implement.

In this work, we propose the perceptual font manifolds from a generative model based on a deep learning approach [5]. For simplicity, we only focus on the capital letter "A" fonts. The results of the perceptual studies in the latent space of the generative model were visualized. In the font generation, we utilize the VAE network for font generation. As an advantage of the VAE model, the latent space is continuous such that similar images are classified close to each other; hence, enabling the exploring of the perception distribution of fonts. In terms of the generation results from VAE model, we propose a user interface for a perceptual study of font styles. When the user changes the latent variables in the multi-dimensional space from the learning results, the user interface displays the corresponding font image as an output. In order to investigate perceptions of font images, we surveyed using three perceptions of casual, formal, and POP font styles. Finally, we propose a font exploring user interface based on the proposed perceptual font manifolds. The proposed user interface is shown to be efficient and easy to use in contrast to other widely used font exploring interfaces.

The main contributions of this research are as follows: (1) we propose a perceptual font manifold to meet specific perceptual requirement in font exploration and editing; (2) we provide a font-exploring user interface based on our proposed perceptual font manifolds, which is verified to be efficient and user-friendly.

## 2. Related Works

In this section, we will discuss the related work about generative models using deep learning approaches, design user interface for graphical applications, and font generation and exploration.

### 2.1 Generative Models

Recently deep neural network has achieved fruitful progresses in machine learning and computer vision fields. Among these network structures, generative models can create photo realistic images learning from huge datasets which are useful in graphics applications, such as VAEs [1] and GANs [2]. Based on these achievements, pix2pix proposed the conditional adversarial networks for image-to-image translation issues [6]. In computer graphics community, the similar networks have been adopted in terrain modeling [7] and flow design [8] problems with paired sketch and image datasets. There are also some interesting work to generate Chinese font using the conditional GAN networks, zi2zi is trained to learn the Chinese character style with paired character images as training dataset [9]. Besides of the rapid development of generative models, it is still challenging to understand the human perception in neural networks.

### 2.2 Design User Interfaces

The research of user interface design plays an important role in computer graphics and human computer interaction. In our daily creativity activities such as painting and sculpting, the commercial tools are too complicated and time-consuming for common users. To solve these issues, the interactive design system can be developed to automatically complete the tedious repetition in painting processes [10], and fulfill the physical functions in complicated functional designs [11]. With the recognition of user gestures and postures, a painting system was engaged in live presentation for augmenting the storytelling performance [12]. A spatial user interface can be used for sculpting and large-scale fabrication with the help of projection mapping techniques [13, 14]. In this work, we especially focus on the design interface for generated fonts with perceptual manifolds.

### 2.3 Font Generation and Exploration

To generate new font style from existed font examples, Paul et al. proposed a neural network architecture, which creates a character font dataset from a single input font image. The dataset contains 62-letter fonts (from "A" to "Z", "a" to "z", and "0" to "9") [15]. An end-to-end system, called Multi-Content GAN, which generates a set of font images using two networks, glyph network and ornamentation network [16]. This system can transfer the typographic stylization as well as the textual stylization (color gradients and effects). EasyFont was proposed to generate large-scale handwriting fonts using non-rigid point set registration approach [17]. We observed that researches that deal with the combination of a generative model and character fonts focus mainly on the generation of character font images; however, they do not put the perceptions of font images into consideration. Therefore, we try to bridge the gap observed above by considering the perception distribution in the latent space of the generative model.

There are several previous works about font recognition and exploration. DeepFont utilized convolutional neural network approach to automatically identify fonts from an image or photo [18]. A character recognition system was proposed using artificial neural network and nearest neighbour approach from scanned images [19]. Using crowdsourcing to collect font attribute data, Donovan et al. proposed font selection interfaces to predict attribute values for new fonts [20]. An interactive font search method was proposed to suggest font candidates by deforming a displayed font image into user's desired font [21]. FontMatcher examined which character font corresponds to the input image and a large number of character fonts are mapped on the semantic space to combine character fonts with human perception [22]. In this work, we utilized the generative model for mapping to the latent space and considered the feature values of character images.

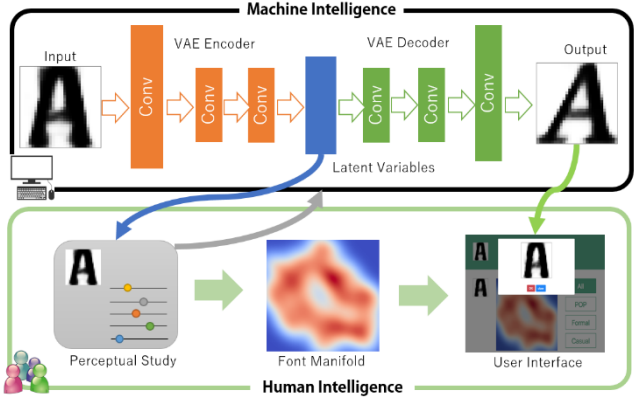

Figure 2. Framework of our proposed perceptual font manifolds. In the machine intelligence process, we employed the VAE generative model obtained the latent space of font generation. A perceptual study in the latent space is explored for three types of perception in human intelligence process. We obtained the perceptual manifolds from the manifold learning approach. Finally, a user can explore the desired fonts on the user interface.

Manifold learning is a useful approach to represent numerous font data. A font manifold was utilized to visualize the distribution of character fonts, and the user can move on this manifold with a mouse and transform the font image continuously [4]. Chinese font manifold was learned to represent the features of font shapes and generate new fonts [23]. Inspired from these projects, we adopt manifold learning on our generated font latent space. However, the pervious works did not consider the human perception from font images.

## 3. System Overview

In this section, we will describe the proposed framework of perceptual fond manifolds and font-exploration user interface, and the technical details will be introduced in next sections.

### 3.1 System Framework

Figure 2 illustrates the framework of our proposed system to combine both machine intelligence and human intelligence processes. Contrary to the conventional generative model based on deep learning approaches, the human effort is involved in the font generation processes in the proposed system. At the machine intelligence stage, we adopt generative network to generate the latent space of various fonts. At the human intelligence stage, we ask the participants to respond to font perception using the perceptual font user interface proposed in this paper. In this user interface, the user can adjust the latent variables to achieve the desired perception. In the case study considered, we recorded all latent variables related to causal, formal, and POP feelings. Furthermore, the perceptual study revealed that the perceptual manifold of fonts is obtained via manifold learning. Finally, we present a user interface for font exploration, which is based on the approaches proposed in this work.

### 3.2 User Interfaces

The proposed font exploring user interface is shown in Figure 3. The proposed user interface is simple and easy to use. The interface shows the corresponding font image on the upper-left window continuously while moving the control point on the heatmap image. To help choose the font in different user perceptions, the user can select the perception buttons on the right side of the user interface (All, POP, Formal, and Casual). The corresponding heatmap representation from a different perspective is presented to the end user. When the user finds the desired font image, the user can click on it, and the modal window will be displayed for further confirmation.

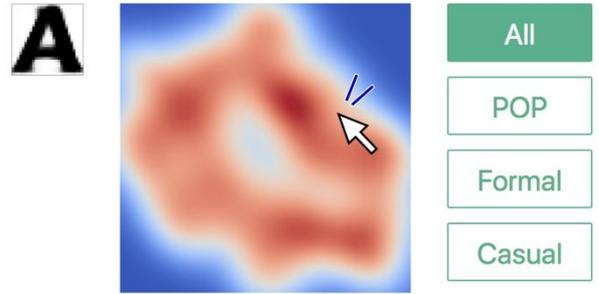

Figure 3. Our proposed user interface with perceptual manifold.

## 4. Font Generation

This work adopts the generative model for font generation. We clarified the learning network; moreover, the details of our dataset are as follows.

### 4.1 Generative Model

In this paper, we utilized the VAE, which is the most representative generative model in the deep learning field [1, 24]. Figure 4 shows the network architecture of the VAE model employed in this paper. The number under each layer indicates the size of each array. The encoder has four convolutional layers, and a flatten layer, while the decoder has a lambda layer and two convolutional layers. The lambda layer calculates the latent variables using a mean and a deviation. For the implementation of the VAE model, we used the open-source neural network library the Keras library on tensorflow [25].

### 4.2 Training Dataset

To construct the training dataset for the VAE network, we downloaded 2244 TrueType font files from Google Fonts. Among all these files, the valid image data includes 2169 PNG image files because 75 files could not be converted into image files. Then, we executed the data cleansing of the font images ready for machine learning.

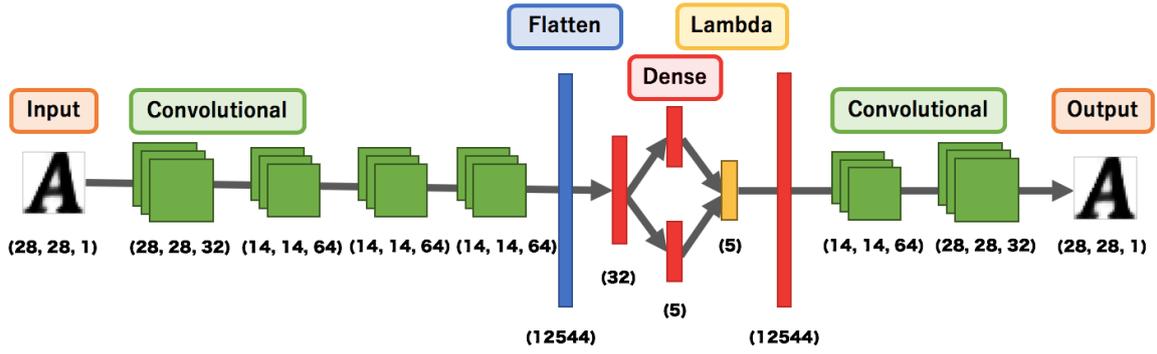

Figure 4. The architecture of our VAE neural network for font generation.

Figure 5 illustrates the data cleansing process. The goal of the data cleansing process is to create 28×28 images for our learning network. Similar to MNIST dataset, we utilized the same learning image sizes. Usually, the training image has a white margin part except for the font body. We erased the white margin of font images and maintained the font body as much as possible for all image data.

In the data cleansing process, we first converted the downloaded TrueType font files into grayscale image files. At the beginning of the process, the image files were set in a resolution with 256×256 pixels and sufficiently large size. Then, the rectangle-bounding box of the font is calculated. To convert the rectangular image into a square size, we added the white row (column) to the shorter side of the rectangle alternately. Finally, we scaled the image to 28×28 pixels using bilinear interpolation. Figure 6 shows examples of font images in our font training dataset.

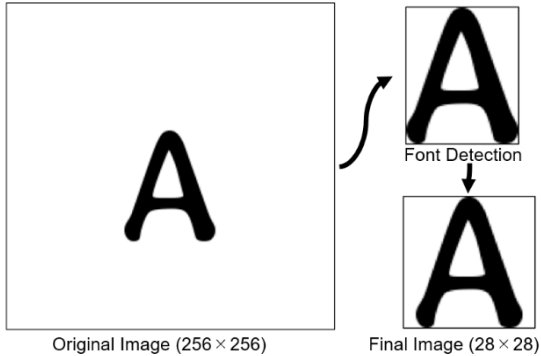

Figure 5. Data cleansing process.

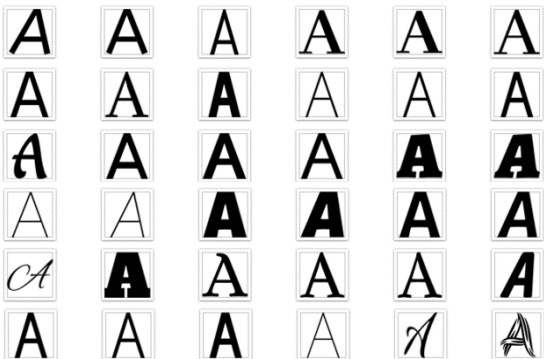

Figure 6. Font examples in our training dataset.

## 5. Perceptual Font Manifold

In this section, we propose the perceptual font manifold considering the user perception in the font generation and exploration. To effectively analyze the user perception of font images, we conducted a perceptual study of font styles based on the generated results from the VAE network. In this study, the participants can choose the latent variables with the output of the font image generated in the latent space. The latent space will have enough information if the latent dimensions are high. However, it is difficult to obtain data from the perceptual study in high-dimensional latent space. Considering the trade-off between the computation cost and the perception evaluation load, in our study, we chose five-dimensional latent space.

### 5.1 Perceptual Study

Figure 7 (a) shows the user interface for the perceptual study. The participants were asked to select their favorite fonts using the five sliders to change the values of the latent variables as shown in Figure 7 (b). Observe that each latent variable can be operated from 0 to 99. However, because the latent space is calculated based on Gaussian distribution, we adopt the percent point function and divide the variable range from 5% to 95% into one hundred equal parts. In our proposed user interface, the output font image corresponding to the selected latent variable is displayed in real time.

To classify the fonts into different user perceptions, we adopted three styles of font perception in this paper, including POP, formal, and casual styles. Whenever the user feels the right perception from the font image, the user was asked to click the perception buttons on the user interface, and the system will save the selected latent variables. To achieve a good starting point, the user can also click the "Changing a Starting Font" button to change all latent variables randomly.

The purpose of this perceptual study is to classify the font images into three designated user perceptions. At the beginning of the perceptual study, we showed examples of feature font of three user perceptions as shown in Figure 8. Besides, we invited 17 graduate students to join our perceptual study. The survey time was limited to five minutes because the user may feel exhausted to repeat the same task for a long time period.

Finally, a total of 884 latent variable sets were collected including 273 POP, 311 formal, and 298 casual styles.

## 5.2 Manifold Learning

The outcome of the perception study of font images revealed that one could obtain distribution data in a latent space of five dimensions. To reduce the distribution data in two dimensions, we adopted a manifold learning approach. In this paper, we utilized the t-distributed Stochastic Neighbor Embedding (tSNE) method for model reduction [26]. Figure 9 shows the result of the dimensionality reduction of the distribution data from our perceptual study. Blue points indicate the POP style, green points denote formal style, and yellow points for causal style of fonts. tSNE-1 and tSNE-2 are the dimensions of the reduced distribution data.

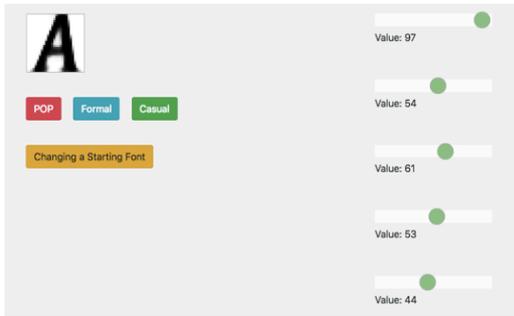

(a) User interface for perceptual study.

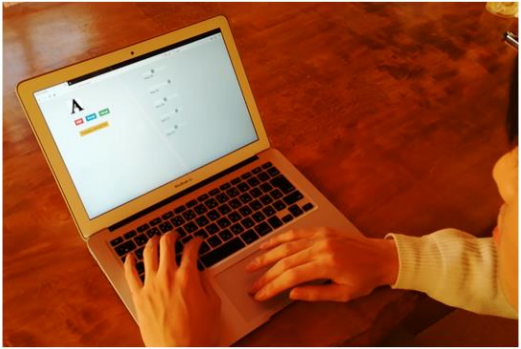

(b) Perceptual study with the proposed UI.

Figure 7. The proposed font selection user interface for perceptual study (a), and one experiment scene in our perceptual study (b).

Figure 8. Feature font examples.

To visualize the two dimensional distribution data from the manifold learning, we adopted the kernel density estimation method to obtain a heatmap representation of the perceptual manifold of fonts [27]. Consequently, we achieved the heatmaps of all perception styles; POP, formal, and casual styles of fonts as shown in Figure 10. The high-density area is displayed in red color, while the low-density area is displayed in blue color. In this work, perceptual manifolds are obtained from the results of a perceptual study of the feature space of fonts. The feature space is learned from the latent space of a generative model of the font generation.

Using the perceptual manifold of fonts introduced in this paper, we propose a user interface for font exploration from the VAE generative network. Figure 3 shows the user interface in our case study, which is implemented as a Web application for easy usage.

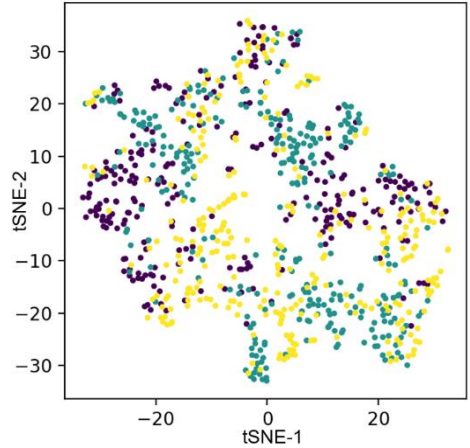

Figure 9. Results of dimensionality reduction from perceptual study.

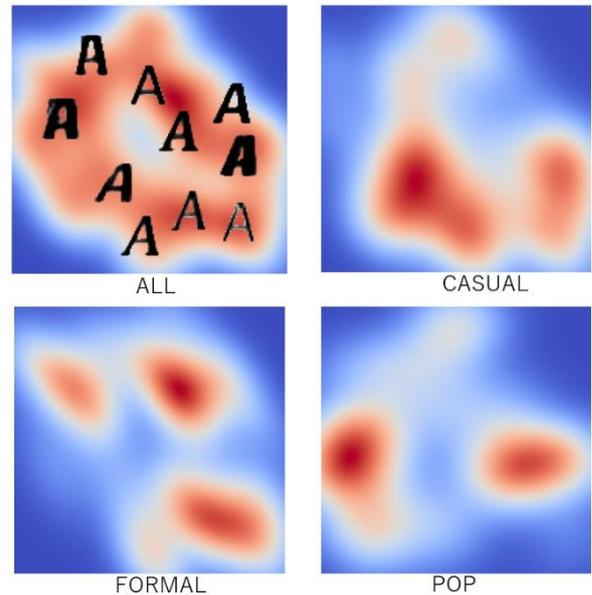

Figure 10. The perceptual manifolds of fonts.

## 6. Case Study

In this section, we present the comparison study of our proposed user interface with conventional font-exploration user interface. For a quantitative analysis, we compared the exploration accuracy and time cost of two interfaces.

## 6.1 Comparison Study

In our case study, we compared the user interface proposed here with the traditional user interface for font searching as used in online font libraries such as Font Squirrel. Note that we only focused on "A" font in this research for simplicity purpose. Figure 11 shows the conventional user interface for font exploration, which displays 1592 font images from the generation results of the VAE network. All these font images are arranged in 10 columns and 160 rows with the scrollbar in a web application.

We randomly selected 10 target font images from the generated results as shown in Figure 12. In both the conventional and the proposed user interfaces, the target font images are located on the upper-left side of the interface windows. The participants were asked to look for the target font images on both user interfaces and click on the explored position with a confirmation window. Figure 12 shows the distribution of target images on the proposed user interface.

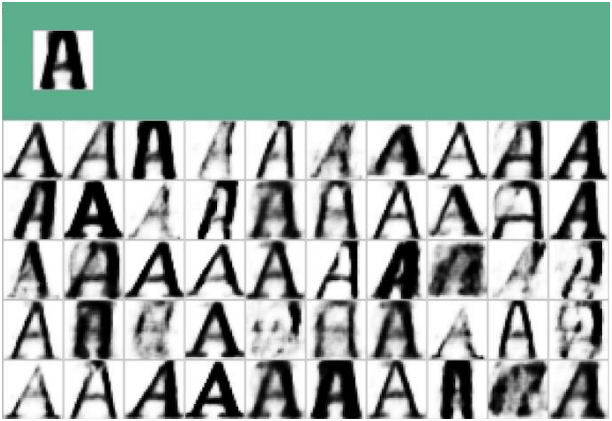

Figure 11. Conventional user interface for font exploration in our comparison study.

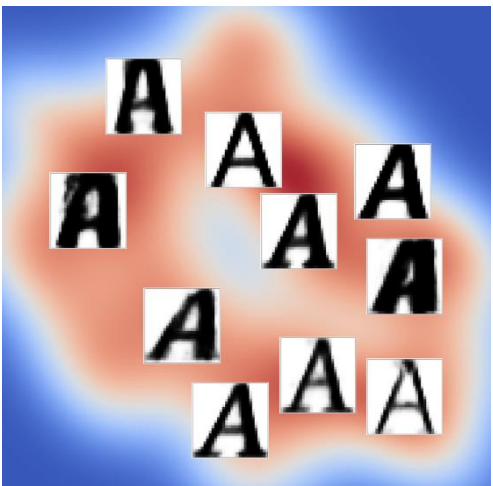

Figure 12. Test dataset used in our case study and their distributions in the manifold of fonts.

We conducted the comparison study with 20 graduate students and randomly divided them into two groups; group 1 and group 2. The members in group 1 were asked to use the conventional user interface first, and after that, to use the proposed interface (Figure 3). The members in group 2 were asked to use the proposed user interface first, and then the conventional user interface.

## 6.2 Discussion

In the case study carried out in this paper, regarding the usage of the two user interfaces considered here, we compared the exploration accuracy and time cost. Regarding the exploration accuracy, we used structural similarity (SSIM) to quantify the similarity between two images [28]. In our research, SSIM scores are calculated between the target font images and the explored images from user operation. The formulation of SSIM score is given as follows.

$$SSIM(x,y) = \frac{(2\mu_x\mu_y + C_1)(2\sigma_{xy} + C_2)}{(\mu_x^2 + \mu_y^2 + C_1)(\sigma_x^2 + \sigma_y^2 + C_2)} \quad (1)$$

where x and y are pixel positions on two font images. $\mu_x$ and $\mu_y$ denote mean pixel values, $\sigma_x$ and $\sigma_y$ are standard deviations of pixel value on two images, and $\sigma_{xy}$ is the covariance value. C1 = 6.5 and C2 = 58.5 are constant values. SSIM score is expressed between 0 and 1, and the closer to 1, the higher the similarity.

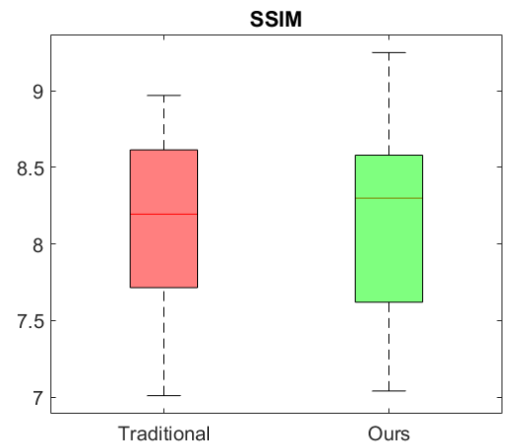

Figure 13. Comparison of SSIM scores between traditional (red) and our proposed user interface (green).

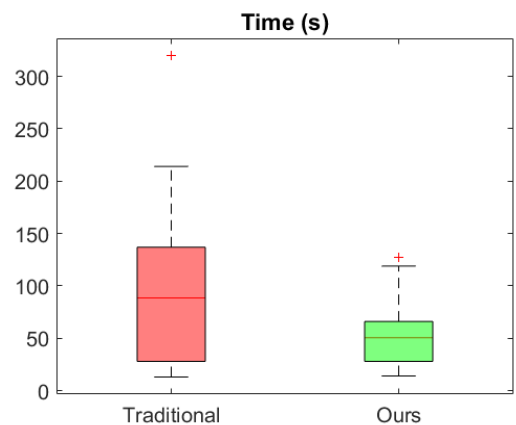

Figure 14. Comparison of time usage between traditional (red) and our proposed user interface (green).

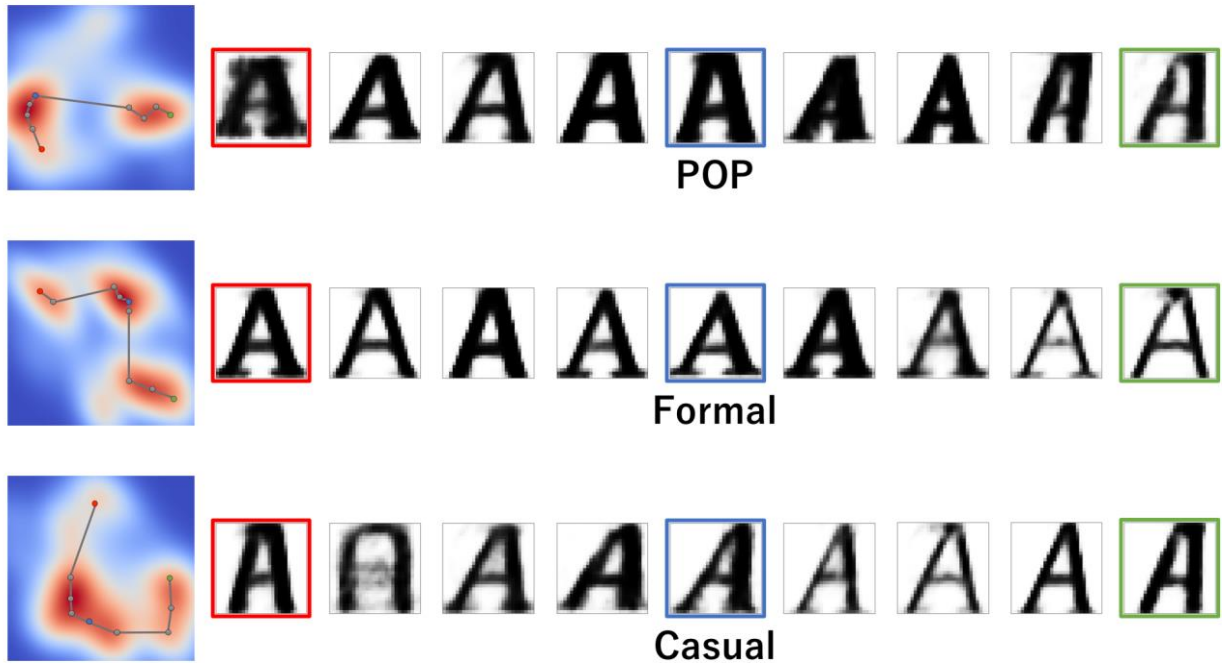

Figure 15. Explorations of font manifolds with three perception style demonstrating our matching results.

Figure 13 shows the comparison results of exploration accuracy in SSIM score using the traditional and our proposed user interface. Regarding SSIM scores, there is no apparent difference between the two interfaces. In details, the user interface proposed in this paper achieved a little higher median and maximum values in SSIM scores than the conventional user interface.

Figure 14 shows the comparison results of time cost in seconds in using the traditional and our proposed user interface. Regarding the average of times cost, the participants operated 1.7 times slower in the traditional interface (97 seconds) than in our proposed interface (54 seconds). To confirm the difference between the two interfaces, we utilized the student's t-test to verify the time cost. The result revealed a score of 0.71%, which shows that there is a significant difference between the two interfaces. Hence, the proposed user interface is more efficient than the conventional user interface in this work.

## 7. Results

The proposed system was implemented on a desktop computer with Ubuntu 16.04 LTS, Intel Core i7-7700 CPU @ 3.60GHz 8 cores, GeForce GTX 1060 3GB GPU, and programmed in Python 3.5 with CUDA 8.0, CuDNN 6.0 and KERAS-gpu 2.1.6. The utilized VAE model was trained for 50 epochs.

Figure 15 shows the exploration results of fonts in three perceptual font manifolds. For POP perception of font images, it is clear that these fonts are in bold and slant styles, whereas they are all in slim shape without slant styles in the formal perception. Also, for casual perception, there is an apparent feature of circular curves in the font styles. All these exploration results agree well with the common sense of these user perceptions as shown in Figure 8 (examples of images).

Though the perception study conducted in this research only handles "A" fonts for simplicity, we can map the explored fonts into all character fonts. Figure 16 shows the text application of our results from "A" to "Z" which are obtained by matching the closest SSIM scores of "A" fonts in existing fonts to our generated font manifolds. The star points on the left figures of the font manifolds denote the font positions. We observed that the perception of the three font styles is apparently correct for common users.

## 8. Conclusion

In this paper, we proposed the perceptual font manifolds using latent space from a generative model and the perceptual study, as a combination of machine intelligence and human intelligence stages. For machine intelligence stage, we employed the generative model of VAE network architecture with five latent dimensions of "A" dataset, whereas for the human intelligence stage, we obtained the distribution data of POP, formal, and casual styles of a font from the latent space in perceptual study. We reduced the data dimensions into two dimensions using manifold learning and made the heat maps of each distribution using kernel density estimation method. Finally, we proposed a user interface for font exploration using our perceptual font manifolds.

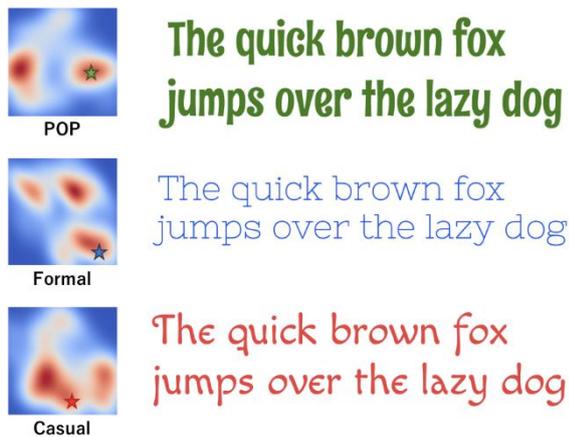

Figure 16. Application of our results with mapping from A to Z.

In our user study, the participants were asked to explore the target font images using the traditional font exploration interface and the user interface proposed in this paper. It was verified that our proposed user interface requires less time cost while keeping good exploration accuracy. Furthermore, the exploration results of different perception styles revealed that our proposed system could achieve the desired style of fonts suitable for not only "A" font but also all character fonts.

As limitations of this work, the proposed system only handles "A" fonts in current research step. Though we have matched all the fonts by finding the closest position in "A" font, the accuracy for other character fonts may be lost. In order to address this issue, the supervised font style transfer employed in [15] should be utilized. Besides, there are many fonts defined in the shapes of outline fonts represented by TrueType font. The outline font expresses the font data as the curve parameters and responds appropriately to scaling. We also considered the exploration of the perceptual manifold of the outline fonts.

As possible future research, the concept of the perceptual study of generative models may be applied to other research targets such as human faces and fashions.

## Reference

[1] Diederik P. Kingma, Max Welling. Auto-Encoding Variational Bays, In Proceeding of the International Conference on Learning Representations (ICLR), 2014.

[2] Ian J. Goodfellow, Jean Pouget-Abadie, Mehdi Mirza, Bing Xu, David Warde-Farley, Sherjil Ozair, Aaron Courville, and Yoshua Bengio. Generative Adversarial Nets. In Proceedings of the 27th International Conference on Neural Information Processing Systems - Volume 2 (NIPS'14), 2014.

[3] Carter Shan and Nielsen Michael. Using Artificial Intelligence to Augment Human Intelligence, Distill, 2017.

[4] Neill D.F. Campbell, Jan Kautz. Learning a Manifold of Fonts, ACM Transactions on Graphics, Vol 33, 4, 2014.

[5] Yuki Fujita, Haoran Xie and Kazunori Miyata. Perceptual Font Manifold from Generative Model, NICOGRAPH International 2019, 2019.

[6] Phillip Isola, Jun-Yan Zhu, Tinghui Zhou, Alexei Efros. Image-to-Image Translation with Conditional Adversarial Networks, arXiv 1611.07004, 2016.

[7] Eric Guérin, Julie Digne, Eric Galin, Adrien Peytavie, Christian Wolf, Bedrich Benes, Benoît Martinez. Interactive Example-Based Terrain Authoring with Conditional Generative Adversarial Networks, ACM Transactions on Graphics, Vol. 36, No. 6, 2017.

[8] Zhongyuan Hu, Haoran Xie, Tsukasa Fukusato, Takahiro Sato and Takeo Igarashi. Sketch2VF: Sketch-based Flow Design with Conditional Generative Adversarial Network. Computer Animation Virtual Worlds, Vol 30, e1889, 2019.

[9] Y. Tian. zi2zi: Master Chinese Calligraphy with Conditional Adversarial Networks. https://github.com/kaonashi-tyc/zi2zi, 2017.

[10] Jun Xing, Hsiang-Ting Chen, and Li-Yi Wei. Autocomplete painting repetitions. ACM Transactions on Graphics. Vol 33, 6, Article 172, 2014.

[11] Haoran Xie, Takeo Igarashi, and Kazunori Miyata. Precomputed Panel Solver for Aerodynamics Simulation. ACM Transactions on Graphics, Vol 37, 2, 2018.

[12] Nazmus Saquib, Rubaiat Habib Kazi, Li-Yi Wei, and Wilmot Li. Interactive Body-Driven Graphics for Augmented Video Performance. CHI Conference on Human Factors in Computing Systems (CHI '19). ACM, Paper 622, 2019.

[13] Alec Rivers, Andrew Adams, and Frado Durand. Sculpting by numbers. ACM Transactions on Graphics. Vol 31, 6, 157, 2012.

[14] Haoran Xie, Yichen Peng, Naiyun Chen, Dazhao Xie, Chia-Ming Chang, and Kazunori Miyata. BalloonFAB: Digital Fabrication of Large-Scale Balloon Art. CHI Conference on Human Factors in Computing Systems (CHI EA '19), 2019.

[15] Paul Upchurch, Noah Snavely, and Kavita Bala. From A to Z: Supervised Transfer of Style and Content Using Deep Neural Network Generators, arXiv: 1603.02003, 2016.

[16] Samaneh Azadi, Matthew Fisher, Vladimir Kim, Zhaowen Wang, Eli Shechtman, and Trevor Darrell. Multi-Content GAN for Few-Shot Font Style Transfer, arXiv: 1712.00516, 2017.

[17] Zhouhui Lian, Bo Zhao, Xudong Chen, and Jianguo Xiao. EasyFont: A Style Learning-Based System to Easily Build Your Large-Scale Handwriting Fonts. ACM Transactions on Graphics. Vol 38, 1, Article 6, 2018.

[18] Wang, Z., Yang, J., Jin, H., Shechtman, E., Agarwala, A., Brandt, J., & Huang, T. S. DeepFont: Identify your font from an image. ACM Multimedia Conference, pp. 451-459, 2015.

[19] H. Mehta, S. Singla and A. Mahajan. Optical character recognition (OCR) system for Roman script English language using Artificial Neural Network (ANN) classifier, International Conference on Research Advances in Integrated Navigation Systems (RAINS), Bangalore, 2016.

[20] Peter Donovan, Janis Libeks, Aseem Agarwala, Aaron Hertzmann. Exploratory Font Selection Using Crowdsourced Attributes. ACM Transactions on Graphics, Vol 33, 4, 2014.

[21] Ken Ishibashi and Kazunori Miyata. Edit-Based Font Search. 22nd International Conference on MultiMedia Modeling (MMM 2016), Vol. 9516. Springer-Verlag, 550-561, 2016.

[22] Saemi Choi, Kiyoharu Aizawa, and Nicu sebe. FontMatcher: Font Image Paring for Harmonious Digital Graphic Design, 23rd International Conference on Intelligent User Interfaces (IUI '18), pp.37-41, 2018.

[23] Yuan Guo, Zhouhui Lian, Yungmin Tang, and Jianguo Xiao.

Creating New Chinese Fonts Based on Manifold Learning and Adversarial Networks, EUROGRAPHICS, Short Papers, 61-64, 2018.

[24] Danilo J. Rezende, Shakir Mohamed, and Daan Wierstra. Stochastic Back-propagation and Approximate Inference in Deep Generative Models, Technical report, arXiv: 1401.4082. 2014.

[25] Francois Chollet. Deep Learning with Python. Manning Publications, 2017.

[26] Laurens van der Maaten, Geofferey Hinton. Visualizing Data using t-SNE, Journal of machine Learning Research, Vol 9, pp.2579-2605, 2008.

[27] B. W. Silverman. Density estimation for statistics and data analysis, Chapman and Hall/CRC, Springer, 1986.

[28] Zhou Wang, Alan Conrad Bovik, Hamid Rahim Sheikh and Euro P. Simoncelli. Image Quality Assessment: From Error Visibility to Structural Similarity, IEEE Transaction on Image Processing, Vol13, 4, 2004.